\documentclass[referee]{raa}           

\usepackage{graphicx,times}
\usepackage{natbib}
\usepackage{amssymb,amsmath}
\usepackage{siunitx}
\bibpunct{(}{)}{;}{a}{}{,}

\newcommand{\CO}[3]{\ensuremath{{}^{#1}\mathrm{C}{}^{#2}{\mathrm{O}}}\ensuremath{#3}}

\newcommand{\vlsr}{\ensuremath{V_{\mathrm{lsr}}}}

\newcommand{\T}[2]{\ensuremath{T_{\mathrm{#1}}^{#2}}}

\newcommand{\nshare}{\ensuremath{N_{\mathrm{share}}}}
\newcommand{\nsame}{\ensuremath{N_{\mathrm{same}}}}

\newcommand{\vcvcen}{\ensuremath{\hat{v}}}
\newcommand{\vcveldisp}{\ensuremath{\hat{\sigma}}}
\newcommand{\vcintint}{\ensuremath{w}}
\newcommand{\vcamp}{\ensuremath{\hat{a}}}

\newcommand{\vcgvcen}{\ensuremath{\hat{V}}}
\newcommand{\vcgveldisp}{\ensuremath{\tilde{\sigma}}}
\newcommand{\vcginttot}{\ensuremath{W}}

\usepackage[pagebackref=true, colorlinks=true, linkcolor=blue, citecolor=blue, filecolor=blue, urlcolor=blue]{hyperref}

\begin{document}

   \title{
   ISMGCC: 
   Finding Gas Structures in Molecular Interstellar Medium Using Gaussian Decomposition and Graph Theory
    }

 \volnopage{ {\bf 20XX} Vol.\ {\bf X} No. {\bf XX}, 000--000}
   \setcounter{page}{1}

   \author{
    Haoran Feng \inst{1,2},
    Zhiwei Chen \inst{1},
    Zhibo Jiang \inst{1,2,3},
    and
    James S. Urquhart \inst{4}
   }
   \institute{ 
    Purple Mountain Observatory, Chinese Academy of Sciences, Nanjing 210023, China; {\it zwchen@pmo.ac.cn}\\
    \and
    University of Science and Technology of China, Chinese Academy of Sciences, Hefei 230026, China \\
    \and
    Center for Astronomy and Space Sciences, Three Gorges University, Yichang 443002, China\\
    \and 
    Centre for Astrophysics and Planetary Science, University of Kent, Canterbury CT2 7NH, UK
\vs \no
   {\small Received 20XX Month Day; accepted 20XX Month Day}
}

\abstract{
   Molecular line emissions are commonly used to trace the distribution and properties of molecular Interstellar Medium (ISM).
   However, the emissions are heavily blended on the Galactic disk toward the inner Galaxy because of the relatively large line widths and the velocity overlaps of spiral arms.
   Structure identification methods based on voxel connectivity in PPV data cubes often produce unrealistically large structures, which is the ``over-linking'' problem.
   Therefore, identifying molecular cloud structures in these directions is not trivial.
   We propose a new method based on Gaussian decomposition and graph theory to solve the over-linking problem, named \texttt{ISMGCC} (InterStellar Medium Gaussian Component Clustering).
   Using the MWISP \CO{13}{}{~(1-0)} data in the range of $13.5^{\circ} \leq l \leq 14.5^{\circ}, |b| \leq 0.5^{\circ}$, and $-100\leq \vlsr \leq +200~\mathrm{km~s^{-1}}$, 
   our method identified three hundred molecular gas structures with at least 16 pixels.
   These structures contain $92\%$ of the total flux in the raw data cube and  show single-peaked line profiles on more than $93\%$ of their pixels.
   The ISMGCC method could distinguish gas structures in crowded regions and retain most of the flux without global data clipping or assumptions on the structure geometry, meanwhile, allowing multiple Gaussian components for complicated line profiles. 
}

   \authorrunning{H. Feng et al. }            %author_head in even pages
   \titlerunning{ISMGCC: Finding Gas Structures in molecular ISM}  % title_head in odd pages
   \maketitle

%________________________________________________ sections below
% 
\section{Introduction}           %% first-level sections will be auto-capitalized
\label{sect:intro}
As the ingredient of star formation, molecular interstellar medium (ISM) is an essential part of the galactic material cycle and plays an indispensable role in the evolution of galaxies.
Observations of molecular line emissions, especially those from carbon monoxide (CO) molecules \citep{2015ARA&A..53..583H}, are commonly used to trace the spatial distribution of molecular gas. 
The radial velocity of molecular lines helps analyze the kinematics and dynamics of molecular gas.
Mapping observations of lines result in data cubes with three dimensions, commonly referred to as PPV (Position-Position-Velocity) data cubes.
However, identifying molecular gas structures in PPV data cubes for further analysis is not trivial.
While manual segmentation is possible when the spatial and velocity coverage is small, many automatic methods have been proposed to fulfill this task on large datasets and specialized for different targets. 
For instance, the algorithms CPROPS \citep{2006PASP..118..590R}, SCIMES \citep{2015MNRAS.454.2067C}, and DBSCAN \citep{DBSCAN,2020ApJ...898...80Y} are applied to identify molecular clouds that are continuous in both spatial and velocity extents. 
Some other methods are aimed at the hierarchical structures of molecular clouds, e.g., ASTRODENDRO \citep{2008ApJ...679.1338R}, DENDROFIND \citep{2012A&A...539A.116W}, and QUICKCLUMP \citep{2017ascl.soft04006S}.
To detect molecular cores or clumps within molecular clouds, ClumpFind \citep{1994ApJ...428..693W,2011ascl.soft07014W}, FellWalker \citep{2015A&C....10...22B}, Gaussclump \citep{1990ApJ...356..513S,2014ascl.soft06018S}, and more recent ones like
LDC \citep{2022RAA....22a5003L}, ConBased \citep{2022A&C....4000613J}, FacetClumps \citep{2023ApJS..267...32J}, SS-3D-Clump \citep{2024A&A...683A.104L} are the appropriate algorithms.
There are also algorithms designed to find filamentary structures, e.g., DisPerSE \citep{2011MNRAS.414..350S,2011MNRAS.414..384S} and FilFinder \citep{2015MNRAS.452.3435K,2016ascl.soft08009K}.

Even though lots of methods have been developed to find various kinds of objects from data cubes, there are still unresolved problems.
For instance, \cite{1987ApJ...319..730S} noticed that the \CO{12}{}{~J=(1-0)} emission toward the galactic disk is blended at $3~\mathrm{K}$ level with immense features as large as $\sim 5^{\circ}$ and $\sim 60~\mathrm{km~s^{-1}}$, which is the result of the over-linking problem in PPV data cubes.
When finding structures in a PPV data cube, each voxel above the cutoff level is an element to be clustered. 
The most commonly used methods, e.g. ASTRODENDRO, cluster the voxels based on the friends-of-friends principle, also known as ``single-linkage'', which means two connected voxels will bring all their previous companions into the same cluster without considering their companions' PPV separations. 
When the dip between two blended spectral peaks is not deep enough, the voxels between the two peaks act as a bridge connecting the two structures into a huge one, even though the original structures could have a large velocity difference. 
Such over-linking problem of single-linkage methods is also known as the ``chaining phenomenon''. 
As we will show in Sect.~\ref{sect:data}, the over-linking problem can create clustering results stretching across multiple spiral arms. 
This often happens in CO emission data because of its high intensity and large line width.
Increasing the intensity cutoff might split the enormous structures but at the cost of losing a large portion of flux. 

\cite{2017ApJ...834...57M} designed a hierarchical clustering method with Gaussian decomposition to solve the over-linking problem, but it still needs an intensity threshold decent procedure. 
The method sets a series of integrated intensity levels from large to small and merges the clusters that meet certain criteria above each level. 
Another suitable method is the ACORNS \citep[Agglomerative Clustering for ORganising Nested Structures,][]{2019MNRAS.485.2457H,2020ascl.soft03003H} algorithm.
\citet{2024AJ....167..220Z} utilized ACORNS and confirmed the multi-layer nature of the Cygnus region with distance measurements,
which in turn illustrates the importance of such methods toward velocity-crowded regions.  
The successes of hierarchical clustering \citep{2017ApJ...834...57M}, ACORNS \citep{2019MNRAS.485.2457H}, and a more pioneer method named FIVE \citep[Friends In VElocity,][]{2013A&A...554A..55H} inspired us that decomposing the spectra into Gaussian components is a promising step in solving the problem.
Methods that directly manipulate PPV voxels above certain thresholds do not often consider the inequivalence between the connectivity on spatial and velocity axes,
but these axes can be separately processed after Gaussian decomposition.

In this work, we propose a method that finds gas structures in emission line data cubes.
It is named InterStellar Medium Gaussian Component Clustering (ISMGCC).
ISMGCC is designed to find structures with continuity in both spatial and velocity directions and keep most pixels of each structure having single-peaked line profiles. 
Meanwhile, it recovers most flux from the raw data cube.
This paper is organized as follows:
In Sect.~\ref{sect:method}, we give a detailed description of the ISMGCC method.
Then in Sect. \ref{sect:data}, we describe the dataset used to test ISMGCC.
Section~\ref{sect:results} presents the Gaussian decomposition result, establishes a set of metrics to evaluate the method performance, shows the parameter experiments, and demonstrates the structure identification results.
In Sect.~\ref{sect:discuss}, 
we discuss some important features, connections with the previous methods, and the limitation of ISMGCC.
Finally, the paper is summarized in Section~\ref{sect:summary}.

\section{Method}
\label{sect:method}
Here we propose the ISMGCC method for molecular gas structure identification in PPV data cubes. 
A molecular gas structure is a PPV region that contains significant emissions with continuity in both spatial and velocity dimensions.
The line profile along the velocity axis tends to be single-peaked at each spatial location.
Our method solves the over-linking problem without extra flux loss or assumptions on the structure geometry.
The main idea is to split the structures with soft boundaries based on probability, with which the problem is converted into finding communities in a weighted graph.

One cause of the over-linking problem is the broad widths of emission lines.
The large supersonic line width has been identified since the first detection of \CO{}{}{~J=(1-0)} \citep{1970ApJ...161L..43W}.
Many factors can broaden the line profile, e.g., turbulence and high optical depth.
In crowded regions, spectral peaks from multiple structures can be blended 
due to their small velocity separation compared to their line widths.
To overcome this problem, one can use Gaussian components  to fit the spectrum at each location, which can be considered a transformation from the ``dense'' PPV data cube into sparse points in PPV space.
This transformation can be expressed as
\begin{equation} \label{eq:gaussfit}
    \{\T{MB}{}(x, y, z)\} \Rightarrow \{g(v;\vcamp_i, \vcvcen_i, \vcveldisp_i, x'_i, y'_i) : i \in \mathbb{N}, i\leq N \},
\end{equation}
where $N$ is the number of Gaussian components used to fit the data cube, $i$ is the index of the Gaussian component, and $g(v;\vcamp, \vcvcen, \vcveldisp, x', y')$ is a Gaussian profile as the function of radial velocity $v$,
\begin{equation}
    g(v; \vcamp, \vcvcen, \vcveldisp, x', y') = \vcamp \exp\left[-\frac{(v - \vcvcen)^2}{2\vcveldisp^2}\right].
\end{equation}
The parameters to describe a Gaussian component include amplitude ($\vcamp$), centroid velocity ($\vcvcen$), velocity dispersion $(\vcveldisp)$, and its spatial location $(x', y')$.

All the procedures in our method are based on the results of Gaussian decomposition.
Any Gaussian decomposition algorithm can be used to create the input table.
Many spectral fitting techniques have been developed in previous works, including semi-automatic ones, for example, 
SCOUSE, SCOUSEPY \citep{2016MNRAS.457.2675H,2019MNRAS.485.2457H}, and fully-automatic ones, such as, GaussPy \citep{2015AJ....149..138L}, BTS \citep[Behind The Spectrum,][]{2018MNRAS.479.1722C}, and the one applied in this work, \textsc{GaussPy+} \citep{2019A&A...628A..78R}.
Users of our method can select the most suitable decomposition method for their dataset.
This decoupling brings the potential of ISMGCC to use line profiles other than Gaussian, e.g., Lorentz and Voigt.
As long as the line profile is symmetric and can be described with parameters similar to $\vcamp, \vcvcen,$ and $\vcveldisp$, i.e., height, center, and width, ISMGCC could adapt it with minimal modification. 

ISMGCC is more like a tool in the intermediate stage of finding the objects for scientific analysis.
It cuts the data along the velocity dimension and creates 2D maps for each structure.
A complete workflow can be decomposing the data cube into Gaussian components , then finding structures with ISMGCC, creating 2D maps for each structure, and finally, utilizing other tools on the 2D maps to define the objects of study based on scientific goals. 
For example, users who want to find clumps or filaments can process the data cube with Gaussian decomposition and ISMGCC, and then use some specific methods (e.g. GaussClump, FilFinder)  on the 2D maps generated by ISMGCC to find the objects of their interests. 

As the Gaussian decomposition procedure is not part of ISMGCC, we will start with a collection of Gaussian components as the entry point to describe the method step-by-step. 
Our application of \textsc{GaussPy+} to decompose the data cube is in Sect.~\ref{sect:gauss_decomp}.
The first operation determines the velocity coherence between Gaussian components  (Sect.~\ref{sect:aperture_clustering}), with which a weighted graph can be established using the probability of Gaussian component  pairs belonging to the same structure as edges (Sect.~\ref{sect:build_graph}).
In Sects.~\ref{sect:find_community}~-~\ref{sect:post_process}, we describe how the Gaussian components in a graph are clustered into structures and the definition of their PPV boundaries. 
At the end of this section, the parameters and implementation of ISMGCC will be briefly introduced (Sect.~\ref{sect:parameters_and_implementation}).

\subsection{Aperture Clustering for Velocity Coherence}
\label{sect:aperture_clustering}

Determining the coherence between Gaussian components requires thoughtful operations. 
Here we describe a robust procedure with only three dimensionless parameters to fulfill the task.
The output of the current step is the probability of two Gaussian components coherent with each other in velocity.

\begin{figure}[h!]
   \center{\includegraphics[width=\textwidth]{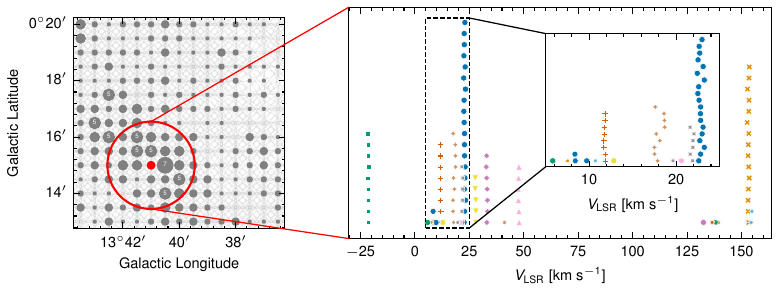}}
   \caption{
      A schematic view of the aperture clustering procedure. 
      The left panel is a field of view where the size of each point is proportional to the number of Gaussian components  on that pixel. 
      Points with more than four Gaussian components are labeled by the number of components therein.
      For each point with at least one Gaussian component, we create an aperture with a radius of $R_{\mathrm{ap}}=3~\mathrm{pix}$, as shown by the light gray circles. 
      The red circle is an instance of such apertures, centered at the red point. 
      The right panel shows the velocity distribution of all Gaussian components in the red circle. 
      Through MeanShift clustering, the Gaussian components are clustered into groups denoted by different colors and shapes.
      In the right panel, Gaussian components in the same group are vertically stacked to show the component number  in each group.
      }
   \label{fig:vcluster}
\end{figure}

\begin{figure}[h!]
   \center{\includegraphics[width=\textwidth]{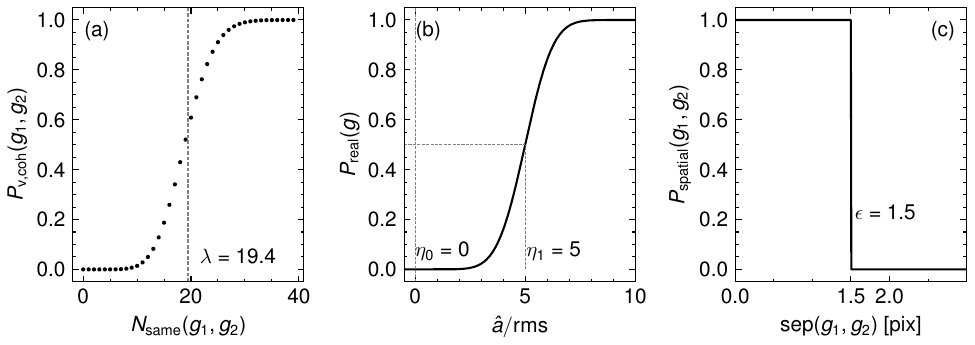}}
   \caption{
      Curves of the functions that help define the probability of Gaussian components  belonging to the same structure. 
      Panel (a) and (b) correspond to Eq. \ref{eq:prob_vcoh} and Eq. \ref{eq:prob_real_signal}, respectively.
      Panel (c) demonstrates the step function that converts the spatial separation between two Gaussian components into a probability of them having spatial connectivity, which corresponds to Eq.~\ref{eq:prob_spatial}.
      }
   \label{fig:prob_func}
\end{figure}

As illustrated by Fig. \ref{fig:vcluster}, circular apertures with radii $R_{\mathrm{ap}} = 3~\mathrm{pix}$ centered at each location with detected Gaussian components  are created.
Gaussian components in each aperture are clustered by their centroid velocity ($\vcvcen$) with the MeanShift algorithm \citep{meanshift}.
This algorithm moves the components toward the local density peak iteratively until the movements stop.
Gaussian components that end up at the same place are clustered as the same group. 
In the right panel of Fig. \ref{fig:vcluster}, we plot the Gaussian components with different colors and markers based on the MeanShift clustering results.
Gaussian components  within the same group are vertically stacked while their horizontal locations are their $\vcvcen$.
Hereafter, we will refer to Gaussian components in the same group in each aperture as a Gaussian Component Group (GCG).
Because of $R_{\mathrm{ap}} > 1~\mathrm{pix}$, one component can appear in multiple apertures.
Therefore, one Gaussian component could belong to multiple GCGs in different apertures and have different companions.
Specifically, given two Gaussian components ($g_1, g_2$) with their spatial separation smaller than $R_{\mathrm{ap}}$, the number of apertures containing both $g_1$ and $g_2$ is $N_{\mathrm{share}}(g_1, g_2)$, while the number of apertures where $g_1$ and $g_2$ belong to the same GCGs is $N_{\mathrm{same}}(g_1, g_2)$.
This configuration can be utilized to define the probability of two Gaussian components coherent with each other in velocity
\begin{equation} \label{eq:prob_vcoh}
P_{\mathrm{v,coh}}(g_1, g_2) =
\begin{cases}
    ~0 & \text{if } \frac{\nsame(g_1, g2)}{\nshare(g_1, g_2)} < \delta;\\
    ~\frac{\Gamma(\nsame(g_1, g_2) + 1, \lambda)}{\nsame(g_1, g_2) !}, & \text{else},
\end{cases}    
\end{equation}
where $\delta$ is a decision boundary that controls the minimal aperture number proportion of $g_1$ and $g_2$ belonging to the same GCGs.
When $\delta = 0.5$ and $(g_1, g_2)$ are not in the same GCGs for half of the shared apertures, we truncate the probability to zero.
Once $(g_1, g_2)$ have passed this criteria, $\nsame(g_1, g_2)$ becomes important.
The cumulative probability function (CDF) of a Poisson distribution is applied to
convert $\nsame$ into a probability, 
where the expected rate of occurrences $\lambda$ can be derived from the mean value of $\nsame$ among all Gaussian component pairs with $N_{\mathrm{share}}(g_1, g_2) > 0$.
Fig. \ref{fig:prob_func}(a) shows an example of the curve between $\nsame$ and $P_{\mathrm{v,coh}}$.

The parameters involved in this step include $R_{\mathrm{ap}}$, $\delta$, $\lambda$, and \texttt{bandwidth} of the MeanShift algorithm.
The expected number of occurrences $\lambda$ can be directly derived from the distribution of $\nsame$.
The \texttt{bandwidth} parameter of the MeanShift algorithm controls the flat kernel width that determines the probability density along the velocity axis, which is in units of $\mathrm{km~s^{-1}}$.
In our practice, we use the amplitude-weighted mean of Gaussian components' velocity dispersion multiplied with a coefficient $\beta$ as the $\texttt{bandwidth}$ of the current aperture
\begin{equation}
    \mathrm{bandwidth} = \beta \frac{\sum \vcamp \vcveldisp}{\sum \vcamp},
\end{equation}
where $\beta$ controls the relative amount.
Smaller $\beta$ values lead to stronger separation between Gaussian components along the velocity axis and more GCGs in each aperture.
Through these simplifications, 
only three dimensionless parameters are retained at this stage:
$R_{\mathrm{ap}}$ in units of pixels, $\delta$, and $\beta$.
In Sect. \ref{sect:metrics}, we will explore the parameter space and select an optimized parameter set for usage.

\subsection{Building Weighted Graph}
\label{sect:build_graph}

A molecular gas structure with Gaussian components as its elements can be considered a weighted graph $G$.
The Gaussian components are nodes in $G$, while the edges connect the nodes.
The probabilities of node pairs belonging to the same structure could be the weight of each edge, denoted as $P(g_1, g_2)$.
In the above subsection, we have derived the probability of two Gaussian components coherent in velocity.
To have a more comprehensive definition of $P(g_1, g_2)$,
the spatial connectivity between $g_1, g_2$ and the prominence of each Gaussian components should also be considered.

For spatial connectivity, we use $\epsilon = 1.5~\mathrm{pix}$ as the threshold for determining whether two Gaussian components are spatially connected, as shown in panel (c) of Fig.~\ref{fig:prob_func}. 
The probability of two Gaussian components ($g_1, g_2$) spatially connected is 
\begin{equation} \label{eq:prob_spatial}
    P_\mathrm{spatial}(g_1, g_2) = \begin{cases}
        ~0, & \mathrm{sep}(g_1, g_2) \geq \epsilon; \\
        ~1, &  \mathrm{sep}(g_1, g_2) < \epsilon,
    \end{cases}
\end{equation}
where $\mathrm{sep}(g_1, g_2)$ is the spatial separation between $g_1$ and $g_2$ in units of pixels. 

We define the probability of a Gaussian component being a real one as 
\begin{equation} \label{eq:prob_real_signal}
    P_\mathrm{real}(g) = \frac{1}{2}\left[1 + \mathrm{erf}\left(\frac{s-\eta_1}{\sqrt{2}}\right)   \right]  \cdot \left[\frac{\mathrm{sign}\left(s - \eta_0 \right) + 1}{2} \right],
\end{equation}
where $s=\vcamp / \mathrm{rms}$ is analog to the signal-to-noise ratio.
Equation \ref{eq:prob_real_signal} contains two parts. 
The first part with a threshold $\eta_1$ and the error function (erf) is the CDF of a Gaussian distribution, where $\eta_1$ acts as the location parameter for the distribution.
In short, the first part makes $P_{\mathrm{real}} = 0.5$ when $s = \eta_1$.
Meanwhile, the second part with a sign function truncates $P_{\mathrm{real}} = 0$ when $s < \eta_0$.
Panel (b) of Fig.~\ref{fig:prob_func} shows the curve of $P_{\mathrm{real}}(g)$,
revealing that $\eta_0$ sets the minimal $s$ requirement for a Gaussian component to have $P_\mathrm{real} > 0$, while $\eta_1$ controls the value of $s$  when $P_\mathrm{real} = 0.5$.

With all three kinds of probabilities defined, we can define the probability of two Gaussian components belonging to the same structure as
\begin{equation} \label{eq:prob_all}
    P(g_1, g_2) = P_{\mathrm{real}}(g_1) \cdot P_{\mathrm{real}}(g_2) \cdot P_{\mathrm{spatial}}(g_1, g_2) \cdot P_{\mathrm{v,coh}}(g_1, g_2).
\end{equation}
Using all the Gaussian components as the nodes and probability $P$ of Gaussian component pairs as the edge weights, we can build a weighted graph $G$ for further operations.

\subsection{Finding Communities in the Graph}
\label{sect:find_community}

In graph theory,
communities are node groups with higher edge density within groups than between them.
This definition of community allows weak connections between groups without chaining them into one structure.
This feature is suitable to solve the over-linking problem caused by single-linkage clustering.
We have constructed the weighted graph $G$ in the previous subsection, where nodes represent Gaussian components and edge weights denote the probability of Gaussian component pairs belonging to the same structure.
Now the communities in $G$ correspond to the molecular gas structures of our interest.

The first step to find the communities in $G$ is removing all the zero-weight edges where $P(g_1, g_2) = 0$.
This could largely simplify the connections in $G$ and split it into multiple subgraphs $\{G_i\}$.
These subgraphs can be separately processed in parallel.
For every subgraph $G_i$, we decide whether it has already been a good molecular gas structure.
If there is any pair of Gaussian components ($g_1, g_2$) in $G_i$ meets the following criteria: 
(a) $g_1$ and $g_2$ are on the same pixel; 
(b) $g_1$ and $g_2$ are at least once clustered into different GCGs in the aperture clustering procedure (Sect. \ref{sect:aperture_clustering}),
we consider $G_i$ as a velocity complicated one because of the multi-GCG signs therein, which requires further segmentation.
If there is no such multi-GCG sign across the entire spatial coverage of $G_i$,
we consider $G_i$ a well-defined molecular gas structure.

\begin{figure}[t!]
   \center{\includegraphics[width=.8\textwidth]{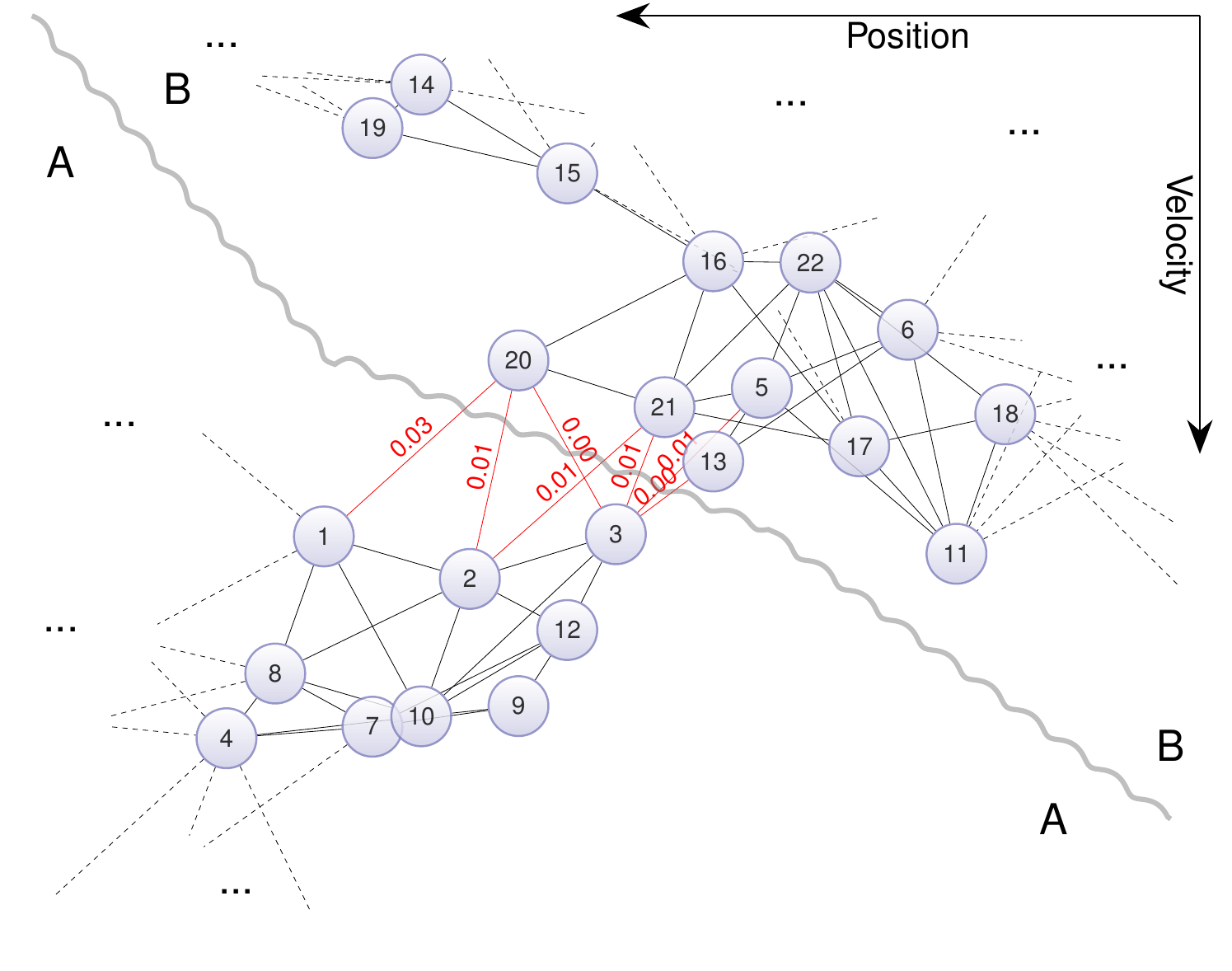}}
   \caption{
      Illustration of finding communities in a subgraph with multiple Gaussian component groups. 
      Each node with a number inside represents a Gaussian component ,
      while the edge connecting two Gaussian components possesses the probability of them belonging to the same structure.
      The coiled line is the boundary determined by the community-finding procedure.
      The average weight value of the solid and dashed black edges is $\sim 0.2$, 
      an order of magnitude larger than the weights of the red edges between Structure A and B.
   }
   \label{fig:graph}
\end{figure}

To visually describe the further segmentation process, we use Fig. \ref{fig:graph} as a schematic diagram to demonstrate the idea.
Each circle with a number inside represents a node in a complicated subgraph.
Locations of the nodes in Fig. \ref{fig:graph} depend on the spatial and velocity coordinates of their corresponding Gaussian components.
The edges between the nodes are represented by the lines connecting the circles, 
where the dashed lines are connected to the nodes omitted for clarity.  
The red edges are labeled with their weights, i.e., the probability of the two connected nodes belonging to the same structure. 
For example, the edge connecting node 1 and node 20 has $P=0.03$ as its weight.
The other red edges between structures A and B also have small weight values.
On the contrary, the other edges in Fig. \ref{fig:graph} have an average weight value of $0.2$, 
an order of magnitude larger than the red ones,
which makes the coiled line in Fig. \ref{fig:graph} a rational boundary between structure A and B.
This probability-based soft boundary is the key to solving the over-linking problem because of its flexibility.
To find such boundaries, we utilize the Clauset-Newman-Moore greedy modularity maximization algorithm \citep{2004PhRvE..70f6111C} to split the subgraph into communities, i.e., the molecular gas structures. 
This algorithm optimizes a quantity named ``modularity'' \citep{2004PhRvE..69b6113N,2004PhRvE..69f6133N}, which measures the quality of a particular graph division. 
It considers each node an individual community at the beginning and keeps combining community pairs until the modularity no longer increases. 
A dimensionless \texttt{resolution} parameter ($\gamma$) required by the algorithm controls the size of communities. 
Smaller $\gamma$ values make the algorithm favor larger community size.
In Sect.~\ref{sect:metrics} we use experiments to select an optimized value of $\gamma$.

\subsection{Post Process}
\label{sect:post_process}

Even though we have determined the Gaussian component members in the molecular gas structures with the previous operations, giving each structure a PPV boundary still helps the recovery of the isolated Gaussian components. 
The isolated Gaussian components are those clustered into structures with $N_\mathrm{pix} < N_{\mathrm{pix,min}}=16$.
The post-process procedure described here would make the valid structures absorb their nearby isolated Gaussian components. 

As will be described in Sect. \ref{sect:gauss_decomp}, due to the complicated profile of the spectral peaks, we allow the existence of multiple Gaussian components for one structure on each pixel.
While the amplitude ($\vcamp$), centroid velocity ($\vcvcen$), and velocity dispersion ($\vcveldisp$) of each Gaussian component are given by \textsc{GaussPy+}, the total integrated intensity ($\vcginttot$), centroid velocity ($\vcgvcen$), and velocity dispersion ($\vcgveldisp$) of the Gaussian component ensemble on a given pixel belonging to the same structure can be derived through the following equations: 
\begin{align}
    \vcginttot & = \sum_{i=1}^{N} \vcintint_i \label{eq:vcginttot}\\
    \vcgvcen & = \frac{\sum_{i=1}^{N} \vcintint_i \vcvcen_i}{\vcginttot}  \label{eq:vcgvcen} \\
    \vcgveldisp & = \left[\frac{\sum_{i=1}^{N} \vcintint_i [\vcveldisp^2 + (\vcvcen_i - \vcgvcen )^2] }{\vcginttot} \right]^{\frac{1}{2}} \label{eq:vcgveldisp}
\end{align}
where $\vcintint$ can be derived from $\vcamp$ and $\vcveldisp$ through
\begin{equation}
    \vcintint = \sqrt{2\pi}\vcamp \vcveldisp.
\end{equation}
Then we can define the upper and lower velocity bound for each pixel of the structure with
$\vcgvcen \pm 1.5\vcgveldisp$.

As for the spatial boundary, the flood fill algorithm implemented in \texttt{scikit-image} \citep{scikit-image} is utilized to define a closed 2D contour on the plane of the sky.
Then we use linear interpolation to fit the upper and lower velocity bounds for each pixel
within this spatial range.
This gives a closed 3D PPV boundary for each structure.
Isolated Gaussian components within the PPV boundary will be absorbed.
Because only the isolated Gaussian components from the invalid structures are reassigned in this step, there would be no Gaussian components belonging to multiple valid structures. 
By now, we have reached the end of the entire method, assigned the decomposed Gaussian components into structures, and given a PPV boundary for each valid structure.

\subsection{Parameters and Implementation} \label{sect:parameters_and_implementation}

There are some parameters involved in the method that control its behavior.
We design the adjustable parameters to be dimensionless or in units of pixels.
These parameters are summarized in Table \ref{tab:parameters}.
Users can change these parameters based on their needs.
As our method can work on the results of any Gaussian decomposition method, the parameters of \texttt{GaussPy+} are not included in Table \ref{tab:parameters}.
In Sect.~\ref{sect:metrics}, we will demonstrate the impact of these parameters with experiments.

\begin{table}[h!]
\center
\caption{Parameters of ISMGCC}
 \begin{tabular}{cccp{0.45\textwidth}}
 \hline
 \hline
  Symbol & Recommended Value & Tested Values & Description\\
 \hline
 $R_{\mathrm{ap}}$      & $3$   & $3, 5, 7$ & 
 Radii of the apertures for velocity coherence clustering, 
 in units of pixels (Sect. \ref{sect:aperture_clustering});
 \\
 $\beta$                &  0.50 & $0.25, 0.50, 0.75, 1.0$ &
 \texttt{bandwidth} coefficient that controls the MeanShift clustering, smaller value leads to more velocity slices (Sect. \ref{sect:aperture_clustering});
 \\
 $\delta$               &  0.5  & $0, 0.5, 0.8$ &
 Minimal value of $\nsame / \nshare$ for two Gaussian components having a chance to be coherent with each other
 (Sect. \ref{sect:aperture_clustering});
 \\
 $\eta_0$               &  0    & $0, 3$ & 
 Amplitude-to-noise threshold for a Gaussian component to have $P_{\mathrm{real}}>0$
 (Eq. \ref{eq:prob_real_signal}, Sect. \ref{sect:build_graph});
 \\
 $\eta_1$               &  5    & $3, 5, 7$ & 
 Amplitude-to-noise value for a Gaussian component to have $P_{\mathrm{real}}=0.5$
 (Eq. \ref{eq:prob_real_signal}, Sect. \ref{sect:build_graph});
 \\
 $\epsilon$             & 1.5   & $1.5$  & 
 Maximum spatial separation between two Gaussian components for being spatially connected, in units of pixels
 (Sect. \ref{sect:build_graph});
 \\
 $\gamma$               &  0.01 & $0.001, 0.01, 0.1, 1.0$ & 
 The \texttt{resolution} parameter of the Clauset-Newman-Moore algorithm \citep{2004PhRvE..70f6111C},
 controls the community sizes in the graph, smaller values lead to larger communities
 (Sect. \ref{sect:find_community});
 \\
 $N_{\mathrm{pix,min}}$ & 16    & $16$   & 
 Minimal Number of pixels for a structure to be considered as a valid one
 (Sect. \ref{sect:post_process}).
 \\
\hline
\end{tabular}
\label{tab:parameters}
\end{table}

ISMGCC is currently implemented with \texttt{Python} and open-source packages.
The manipulations of the weighted graph are through the \texttt{networkx} package \citep{networkx}, including the community finding procedure in Sect.~\ref{sect:find_community} using the \texttt{greedy\_modularity\_communities} function.
We also utilized the MeanShift algorithm \citep{meanshift} implemented in \texttt{scikit-learn} package \citep{scikit-learn} for the aperture clustering procedure in Sect.~\ref{sect:aperture_clustering}.
The post process in Sect.~\ref{sect:post_process} applied \texttt{scikit-image} \citep{scikit-image} to determine the final spatial borders of structures and employed \texttt{scipy} \citep{2020SciPy-NMeth} to interpolate the upper and lower velocity boundaries.
The code is available online\footnote{\url{https://github.com/Haoran-Feng/ismgcc}} with documents and examples.
The output of our method is tables containing the structure IDs of the input Gaussian components .
Functions that convert the result Gaussian component  table into pixel tables and 2D maps will also be provided. 
The detailed usage information will be delivered in the document.

\section{Data}
\label{sect:data}

\begin{figure*}[ht!]
   \center{\includegraphics[width=\textwidth]{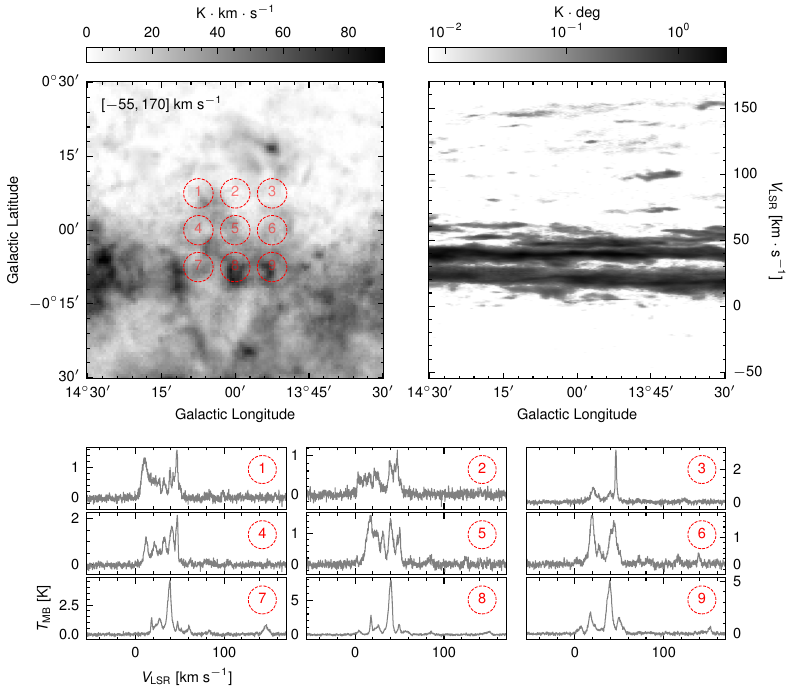}}
   \caption{
      An overview of the MWISP \CO{13}{}{~(1-0)} data used in this work. 
      The upper left panel is the integrated intensity map within the velocity interval of $[-55, 170]~\mathrm{km~s^{-1}}$.
      The upper right panel shows the longitude-velocity map. 
      The nine cells in the bottom contain the average spectra from the nine red apertures shown in the upper left panel.
      }
   \label{fig:data}
\end{figure*}

\begin{figure*}[ht!]
    \centering
    \includegraphics[width=1.0\linewidth]{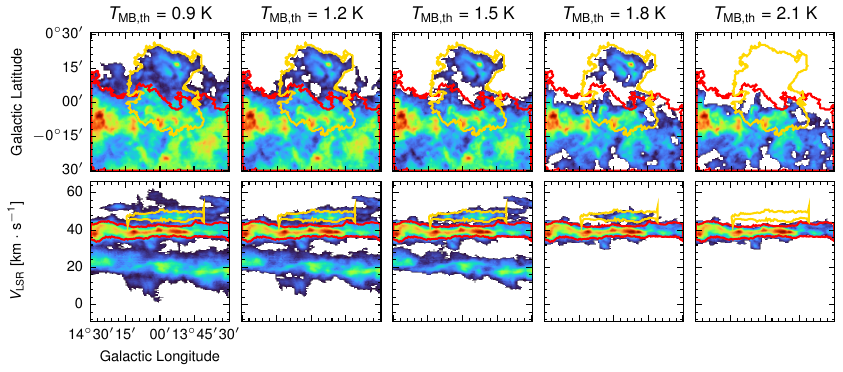}
    \caption{
The largest dendrogram structures above each threshold ($T_{\mathrm{MB, th}}$).  
Each panel column corresponds to a value of $T_{\mathrm{MB}, th}$, where the first row is an integrated intensity map and the second row is the longitude-velocity map. 
The spatial and velocity boundaries of the first two largest structures identified by ISMGCC are denoted as the red and gold contours, respectively.
    }
    \label{fig:dendro-example}
\end{figure*}

We utilize a fraction of the data from the Milky Way Imaging Scroll Painting (MWISP) survey \citep{2019ApJS..240....9S} to test the ISMGCC method.
MWISP observes the entire Galactic plane in the northern sky using the PMO-13.7m telescope. 
Three CO emission lines, \CO{12}{}{}, \CO{13}{}{}, and \CO{}{18}{~J=(1-0)}, are observed simultaneously.
The typical RMS noise levels ($\sigma_{\mathrm{rms}}$) for the three lines are $\sim 0.5, \sim 0.3$, and $\sim 0.3~\mathrm{K}$, respectively, with the corresponding velocity channel widths of $0.158, 0.167, 0.167~\mathrm{km\cdot s^{-1}}$, and beam sizes of $49'', 52'', 52''$.
The observations are taken in position-switch On-The-Fly \citep{2018AcASn..59....3S} mode.
Details on the survey, instruments, and noise analysis can be found in related papers \citep{2019ApJS..240....9S,Yang2008,2012ITTST...2..593S,2021RAA....21..304C}.
From the three lines observed by MWISP, we have chosen \CO{13}{}{~(1-0)} to test our method, because it is optically thin at most locations and has a relatively high detection rate \citep{2023AJ....165..106W,2023AJ....166..121W}.

Spatial and velocity ranges of the utilized MWISP \CO{13}{}{} data are $13.5^{\circ} \leq l \leq 14.5^{\circ}, |b| \leq 0.5^{\circ}$, and $-100 \leq \vlsr \leq +200~\mathrm{km~s^{-1}}$.
The velocity interval is wide enough to enclose most of the emission from the galaxy in this direction.
The direction centered at $l=14^{\circ}, b=0^{\circ}$ is pointing at the inner galaxy, which is suitable to evaluate the performance of our method in crowded regions. 
To give an overview of this data cube, we plot the integrated intensity map and longitude-velocity map in the top panels of Fig. \ref{fig:data}.
The velocity interval for the two maps is narrowed down to $[-55, 170]~\mathrm{km~s^{-1}}$ for clarity. 
To reduce the impact of noise, only voxels with three consecutive channels brighter than $3\sigma_{\mathrm{rms}}$ are included in creating the maps.
But all the values reported in this paper are still using the raw data cube with the full velocity interval of $[-100, +200]~\mathrm{km~s^{-1}}$ without any masking operation.
The nine cells in the bottom panels of Fig. \ref{fig:data} show the average \CO{13}{}{~(1-0)} spectra from the nine red apertures in the upper left panel.
These spectra have complicated line profiles, especially in the velocity range of $[0, 60]~\mathrm{km~s^{-1}}$.
The signals peaked at different velocity could be heavily blended.
Using single-linkage clustering methods to find structures in this kind of data would give an immense structure containing the majority of flux in the entire data cube.

To demonstrate the problem visually, we apply ASTRODENDRO on the data cube and plot the largest dendrogram structures above each intensity level in Fig.~\ref{fig:dendro-example}.
On top of each panel column is the threshold $T_{\mathrm{MB, th}}$ value, while the two rows are the integrated intensity maps and longitude-velocity maps, respectively.
The two spiral-arm-like structures around $20$ and $40~\mathrm{km~s^{-1}}$ could not be distinguished until $T_{\mathrm{MB, th}}\gtrsim 1.8~\mathrm{K} \sim 6\sigma_{\mathrm{rms}}$, but there is still a structure $\sim 47.5~\mathrm{km~s^{-1}}$ attached to the $40~\mathrm{km~s^{-1}}$ arm. 
Increasing $T_{\mathrm{MB, th}}$ to $2.1~\mathrm{K}$ finally removes the $47.5~\mathrm{km~s^{-1}}$ part and isolates the one around $40~\mathrm{km~s^{-1}}$, but the emissions in their outskirts are heavily lost. 
For comparison, we also plot the spatial and velocity boundaries of the first two largest structures identified by ISMGCC as red and gold contours in Fig.~\ref{fig:dendro-example}. 
They correspond well with the $40$ and $47.5~\mathrm{km~s^{-1}}$ portions with the outskirts largely retained. 

\section{Results}
\label{sect:results} 

\subsection{Gaussian Decomposition of Spectra}
\label{sect:gauss_decomp}

We utilize \textsc{GaussPy+} \citep{2019A&A...628A..78R} to fit the entire data cube with Gaussian components.
It automatically estimates the noise level of each spectrum, determines the number of Gaussian components, and fits all spectra with considerations of spatial coherence. 
The default parameters of \textsc{GaussPy+} were applied, while the values of the smoothing parameters $\alpha_1 = 2.18$ and $\alpha_2 = 4.94$ were taken from the test on MWISP data \citep{2020A&A...633A..14R}.
On the current dataset, the Gaussian decomposition retains $98.3\%$ of the total flux in the raw data cube.

\begin{figure}[h!]
   \center{\includegraphics[width=\textwidth]{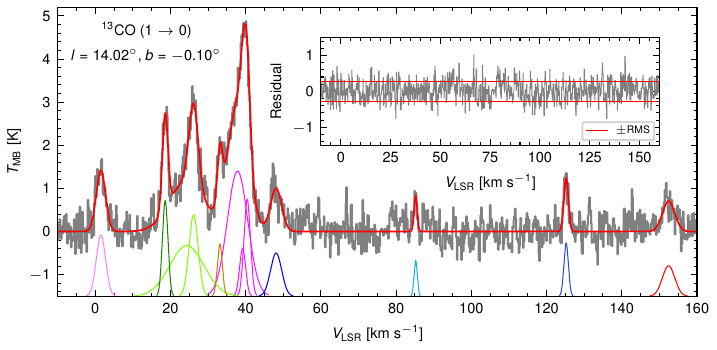}}
   \caption{
      An example of the Gaussian decomposition. 
      This spectrum is taken from the \CO{13}{}{~(1-0)} data cube with pixel size of $30''$ and beam size of $\sim 52''$.
      The Gaussian components  at the bottom are the fitting result of \texttt{gausspy+}
      and colored by the final structure identification result of our method.
      We vertically shifted the Gaussian components by $-1.5~\mathrm{K}$ for clarity. 
      The red curve overlaid on the spectrum is the sum of the Gaussian components.
      The inset panel demonstrates the fitting residual and the RMS noise of $\sim 0.27~\mathrm{K}$. 
      }
   \label{fig:gaussfit}
\end{figure}

Figure~\ref{fig:gaussfit} shows the fitting result centered at $l=14.02^{\circ}, b=-0.10^{\circ}$.
It takes twelve Gaussian components to fit the spectrum. 
However, due to the complexity of the line profiles, multiple Gaussian components are required to describe some of the peaks. 
For example, the highest peak ($\sim 40~\mathrm{km~s^{-1}}$) is decomposed into three Gaussian components because of its skewed shape. 
Another example is the second highest peak ($\sim 25~\mathrm{km~s^{-1}}$) decomposed into two Gaussian components due to its broad base. 
These examples imply that assuming only one Gaussian component exists at each location of a given structure is inappropriate.
Therefore, our method allows multiple Gaussian components on the same pixel to be clustered into the same structure. 
The Gaussian components shown in the bottom of Fig. \ref{fig:gaussfit} are colored by the final clustering result of our method.
It can be seen that our method has a relatively rational segmentation between the structures along this line of sight.

\subsection{Metrics, Experiments, and Parameter Setting} 

\label{sect:metrics}

To evaluate the performance of the ISMGCC method with variable parameter settings, three metrics are applied: Flux Recovery Ratio (FRR), Single Peak Ratio (SPR), and Border Inner Ratio (BIR).
These metrics are designed to reflect performance from three perspectives: flux, velocity coherence, and rationality of spatial borders. 
FRR is the sum of flux in the valid structures divided by the total flux of the data cube.
The valid structures here are those with $N_\mathrm{pix}\geq N_{\mathrm{pix,min}}=16$.

\begin{figure}[t!]
   \center{\includegraphics[width=\textwidth]{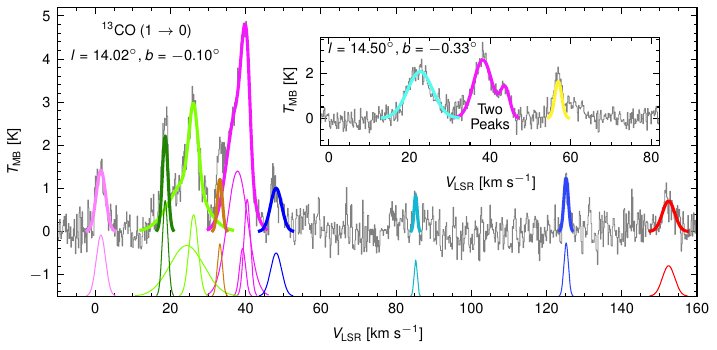}}
   \caption{
      Examples of line profiles of multiple structures. 
      The main panel is the same example as Fig. \ref{fig:gaussfit}. 
      Overlaid on the spectrum is the recovered line profile for each structure.
      The inset panel contains a spectrum from another location, where the structure around $40~\mathrm{km~s^{-1}}$ has a line profile with two peaks. 
      Note that the line profiles in both panels near $40~\mathrm{km~s^{-1}}$ belong to the same structure. 
      }
   \label{fig:two_peaks}
\end{figure}

To evaluate how well the structures are segmented along the velocity axis, we define the SPR as the number of pixels with only one peak divided by the total number of pixels among all structures. 
Specifically, as the Gaussian components have been clustered into structures, we can recover the line profile on each pixel for each structure with the Gaussian components therein.
Using the same spectrum example as Fig. \ref{fig:gaussfit}, we show the recovered line profile of each structure in the main panel of Fig. \ref{fig:two_peaks}.
Even though multiple Gaussian components are sometimes required to describe the line profile, it could still be single-peaked, e.g., the highest one near $40~\mathrm{km~s^{-1}}$ in the main panel of Fig. \ref{fig:two_peaks}.
When there is at least one prominent dip in the line profile, it should be considered a multi-peaked situation, e.g., the one around $40~\mathrm{km~s^{-1}}$ in the inset panel of Fig. \ref{fig:two_peaks}, where the spectrum is taken from another pixel.
A peak must be higher than the dip around it for more than $0.3~\mathrm{K}$, which is close to the RMS noise level of MWISP \CO{13}{}{} data.
Otherwise, it is not considered prominent and will not contribute to the number of peaks.
With the number of peaks on each pixel of each structure derived, 
we can calculate the SPR with
\begin{equation}
   \mathrm{SPR} = 1 -  \frac{\sum_j N_{\mathrm{pix},j}(\mathrm{number~of~peaks} \geq 2)}{\sum_j N_{\mathrm{pix},j}},
\end{equation}
where $j$ is the index of structure, $N_{\mathrm{pix},j}$ is the number of pixels in structure $j$.

Besides the velocity axis, the rationality of structure segmentation along the two spatial dimensions is also important.
Therefore, we use the BIR to reflect the intensity contrast between spatial border pixels and inner structure pixels.
The BIR is the average integrated intensity ratio between the border and inner structure pixels.
The border pixels belong to the structure but touch at least one foreign pixel. 
Pixels from all valid structures are counted together. 
Equation~\ref{eq:vcginttot} defines the integrated intensity of a pixel in a given structure.  

\begin{figure}[t!]
   \center{\includegraphics[width=1.0\textwidth]{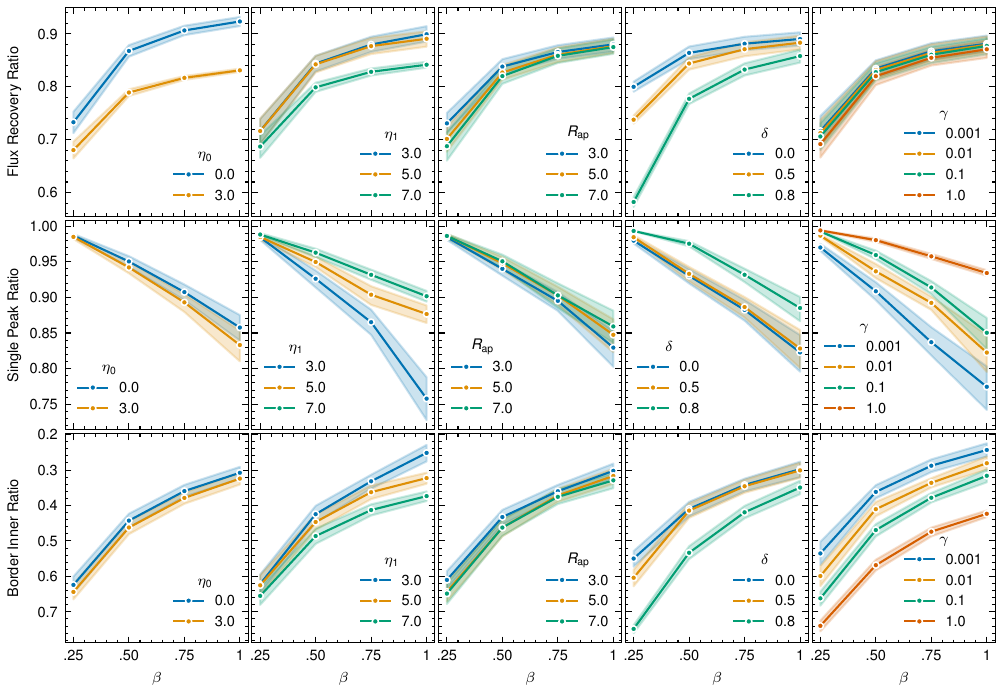}}
   \caption{
      Evaluation of the method performance on three metrics.
      Three rows of panels show the trend of the Flux Recovery Ratio, Single Peak Ratio, and Border Inner Ratio, from top to bottom, respectively.
      All panels use $\beta$, the bandwidth coefficient, as the x-axis.
      The error bands indicate the 95\% confidence interval of the average metric values using bootstrap.
      Note that the y-axes of the bottom row panels are inverted to make the upper side always preferred.
      }
   \label{fig:metrics}
\end{figure}

In consideration of the goal of our method, higher FRR and SPR scores are favored while lower BIR is preferred. 
To evaluate the impact of the parameters shown in Table \ref{tab:parameters} and choose an optimized parameter setting, we tested the method on a parameter grid containing 864 settings with various $\beta, R_{\mathrm{ap}}, \eta_0, \eta_1, \gamma$ values. 
Table \ref{tab:parameters} contains the tested values of each parameter.  
We demonstrate the three metrics of the experiments in Fig. \ref{fig:metrics}.
The FRR, SPR, and BIR as functions of $\beta$ are displayed in the three rows of panels in Fig. \ref{fig:metrics}.
As shown in Table \ref{tab:parameters}, six parameters have more than one tested value.
Therefore, besides $\beta$ on the x-axis, the other parameters are shown as lines with different colors in panel columns, $\eta_0, \eta_1, R_{\mathrm{ap}}, \delta,$ and $\gamma$, from left to right, respectively.
Vertical locations of the points in each panel represent the average metric score with the current $\beta$ and the value of another parameter. 
For example, in the top left panel of Fig. \ref{fig:metrics}, the vertical coordinate of the blue point with $\beta=0.50$ is the average FRR of 108 parameter settings with $\beta=0.50 $ and $\eta_0 = 0$. 
These 108 experiments cover all tested values of the remaining parameters.
The colored error bands around the average values represent the dispersion of the metric values.
The widths of the error bands are derived by the 95\% confidence interval of the average value using bootstrap.

Choosing the setting of parameters is a process of compromise. 
For instance, larger $\beta$ values lead to better FRR and BIR, but damage SPR. 
Therefore, to balance the three metrics, we have chosen $\beta = 0.50$ in the final setting on the current data set. 
Meanwhile, the values of some other parameters are easy to decide.
For example, $\eta_0 = 0.0$ has better FRR than $\eta_0 = 3$ and no obvious adverse impact on SPR and BIR. 
$R_{\mathrm{ap}}$ does not seem to have a large metric difference between the tested values, therefore, we have set it to three for runtime performance.

The final optimized parameter setting is $\beta = 0.5, \eta_0=0, \eta_1=5, R_{\mathrm{ap}}=3, \delta=0.5$, and $\gamma=0.01$.
This setting yields $\mathrm{FRR}=0.92$, $\mathrm{SPR}=0.93$, and $\mathrm{BIR}=0.34$ on the current data set.
With this optimized parameter setting, our method can recover more than $90\%$ of the flux in the raw data cube.
Note that only the flux in structures with $N_{\mathrm{pix}} \geq N_{\mathrm{pix,min}}=16$ 
is included in the numerator. 
Meanwhile, these structures show single-peaked profiles in $93\%$ of their angular area. 

\subsection{Structure Finding Results}

\begin{figure}[t!]
   \center{\includegraphics[width=\textwidth]{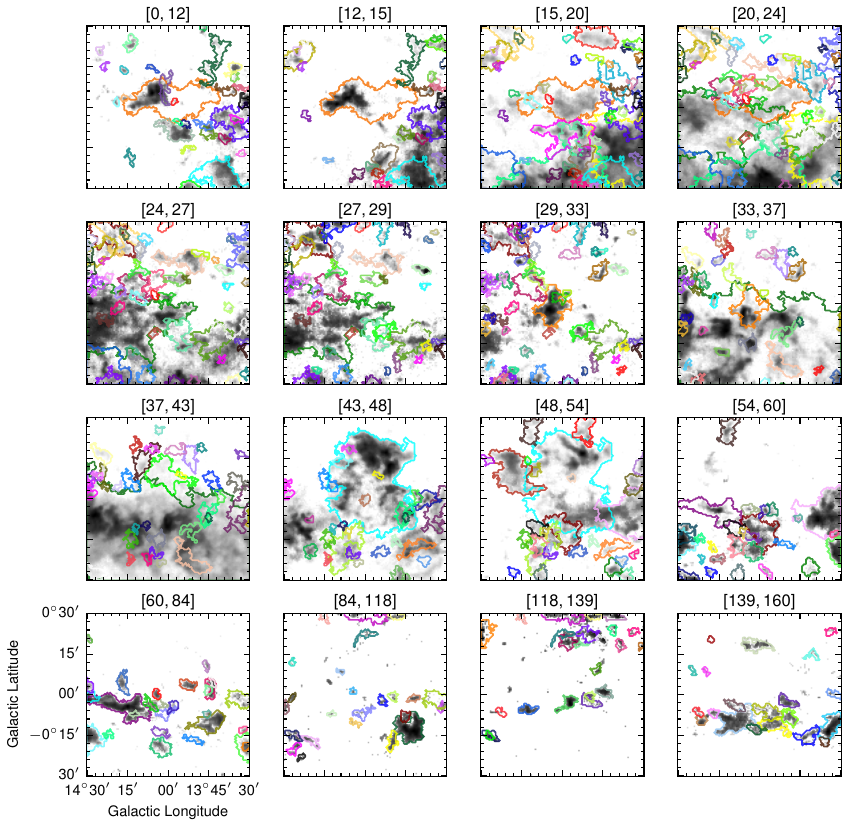}}
   \caption{
      Spatial boundaries of the identified structures overlapped on the integrated intensity maps with various velocity intervals.
      On top of each panel is the velocity interval in units of $\mathrm{km~s^{-1}}$.
      }
   \label{fig:borders_vlayers}
\end{figure}

Our method identified three hundred valid structures in the current dataset with the optimized parameter setting from the previous subsection.
To visually display the structure finding results of our method and demonstrate the spatial segmentation, we draw the spatial boundary of each structure on the integrated intensity maps with various velocity intervals in Fig. \ref{fig:borders_vlayers}.
Because the structures are crowded between $12$ and $60~\mathrm{km~s^{-1}}$, narrower velocity intervals are used in this range to reduce the number of structures in each panel.
Some structures with large velocity spans can appear in multiple panels with the same boundary colors.
For example, the structure outlined by the orange contour appears in the center of all top-row panels, because its large velocity gradient causes a wide velocity span.

\begin{figure}[t!]
   \center{\includegraphics[width=\textwidth]{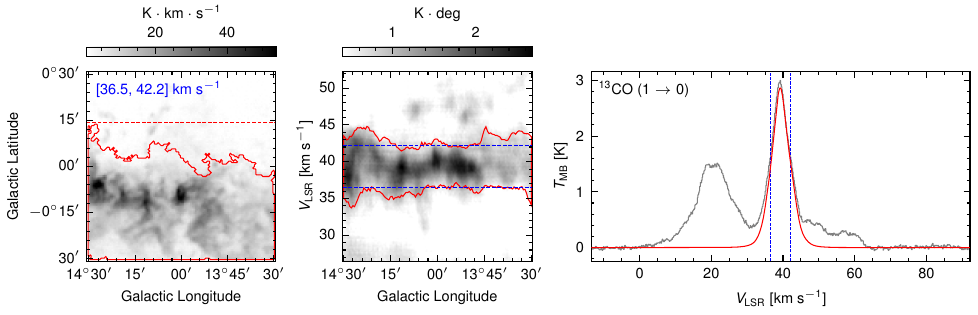}}
   \caption{
      The largest molecular gas structure in the current data set.
      The three panels from left to right are the integrated intensity map, longitude-velocity map, and average spectrum.
      The spectral profile in the right panel is the raw average spectrum within the spatial range denoted by the solid red contour in the left panel.
      The red curve overlapping on the spectrum is the sum of all Gaussian components belonging to the structure. 
      The half-max value of this red curve defines the velocity interval of the integrated intensity map.
      The dashed blue lines in the middle and right panels denote this velocity interval.
      The velocity border in the longitude-velocity map is defined by the median values of $\vcgvcen \pm 1.5\vcgveldisp$ along each longitude, where $\vcgvcen$ and $\vcgveldisp$ are defined by Eq.~\ref{eq:vcgvcen} and Eq.~\ref{eq:vcgveldisp}.
      The dashed red rectangle in the left panel shows the spatial range of the longitude-velocity map.
      }
   \label{fig:struct_example1}
\end{figure}

\begin{figure}[h!]
   \center{\includegraphics[width=\textwidth]{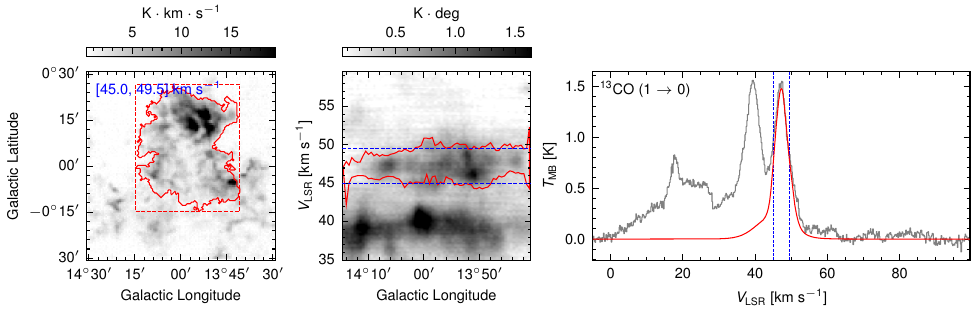}}
   \caption{The second largest structure. Description of the elements is the same as Fig. \ref{fig:struct_example1}.}
   \label{fig:struct_example2}
\end{figure}

In Fig. \ref{fig:borders_vlayers}, one can notice that there is a large structure outlined by dark green contour at the $[33, 37]~\mathrm{km~s^{-1}}$ and $[37, 43]~\mathrm{km~s^{-1}}$ panels.
We plot its integrated intensity map, longitude-velocity map, and average spectrum in Fig. \ref{fig:struct_example1}.
Its average line profile is a single peak around $40~\mathrm{km~s^{-1}}$, while its emission shows continuity in the longitude-velocity map.
Given this velocity information, this structure is well-defined in spatial and velocity dimensions.
In Fig.~\ref{fig:struct_example2}, we show another large-scale structure recovered in the cube. 
Unlike the first example in Fig. \ref{fig:struct_example1} that touches the data edge, this structure has a complete spatial boundary. 
It shows complex internal substructures while the continuity on the velocity dimension is solid.
In the right panel, the recovered average spectrum shown by the red curve perfectly fits the current peak of the raw average spectrum.
The emission corresponds to the other peak around $40~\mathrm{km~s^{-1}}$ can also be seen 
in the longitude-velocity map, which is part of the emission from the first example shown in Fig. \ref{fig:struct_example1}.
The raw average spectrum contains two prominent peaks, one around $40~\mathrm{km~s^{-1}}$ and the other near $47.5~\mathrm{km~s^{-1}}$.
The dip between them is deep but has an intensity value of $\gtrsim 0.6~\mathrm{K}$.
This means that the single-linkage methods processing voxels of the data cube would not be able to disconnect the two structures along the velocity axis unless using a very high cutoff threshold, a consequence of which would be the loss of a significant amount of flux.
The spatial and velocity boundaries of the two structure examples in Fig.~\ref{fig:struct_example1} and Fig.~\ref{fig:struct_example2} are shown as the red and gold contours in Fig.~\ref{fig:dendro-example}.
The method proposed in this work can distinguish emissions from multiple velocity layers without excluding large amounts of diffuse emission.

\begin{figure}[h!]
    \centering
    \includegraphics[width=\textwidth]{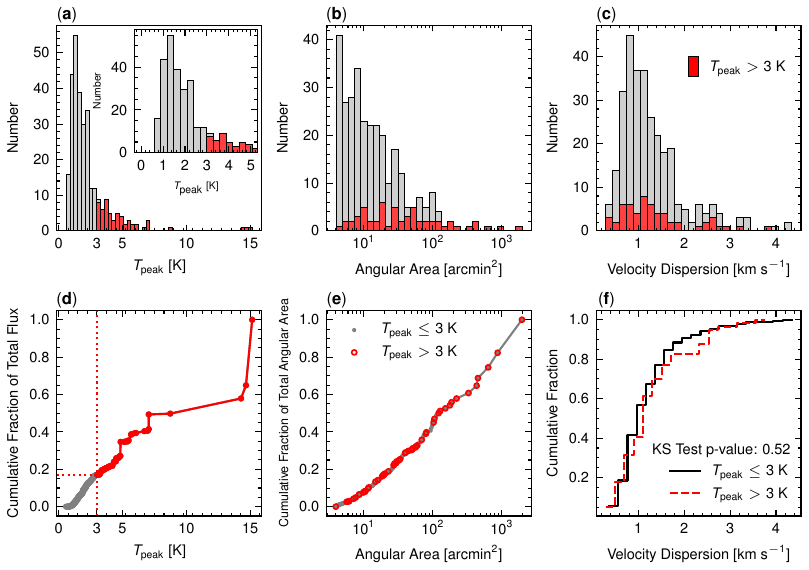}
    \caption{
    Panel (a), (b), and (c) show the distributions of \CO{13}{}{} peak brightness temperature ($T_{\mathrm{peak}}$), angular area, and velocity dispersion of three hundred valid structures identified by ISMGCC.
    The subset with $T_{\mathrm{peak}}>3~\mathrm{K}$ is shown in red in all panels. 
    Panel (d) demonstrates the cumulative fraction of the total flux as a function of structure $T_{\mathrm{peak}}$.
    Panel (e) is similar to panel (d) but for the angular area. 
    Panel (f) presents the velocity dispersion cumulative distributions of the bright and dimmer subsets.
    }
    \label{fig:struct-prop-distrib}
\end{figure}

To reveal the property distributions of the identified structures, we plot the histograms of their \CO{13}{}{} peak brightness temperature ($T_{\mathrm{peak}}$), angular area, and velocity dispersion in Fig.~\ref{fig:struct-prop-distrib} (a), (b), and (c), respectively.
We use the tool delivered with ISMGCC to generate recovered data cubes from clustered Gaussian components to measure these quantities. 
From the distributions shown in Fig.~\ref{fig:struct-prop-distrib}, it can be concluded that most structures have weak \CO{13}{}{} emissions and small angular areas, while a few large structures have high $T_{\mathrm{peak}}$ and possess a large portion of the total flux.
As shown in Fig.~\ref{fig:struct-prop-distrib}(d), the structures with $T_{\mathrm{peak}} \leq 3~\mathrm{K}$ hold only $17\%$ of the total flux. 
The weakest gas structures have $T_{\mathrm{peak}} < 1~\mathrm{K}$, close to three times the RMS noise level.
Given that the reported values are of peaks, the Gaussian decomposition and our method can recover flux from the velocity channels with a signal-to-noise ratio smaller than three. 
Although the flux in the 27 structures with $T_{\mathrm{peak}} < 1~\mathrm{K}$ only constitutes 0.05\% of the total flux in all structures, this weak-signal-sensitive feature of our method could also significantly increase the recovered flux in those relatively large structures. 
For comparison, clipping out the recovered data cubes at $0.3, 0.6,$ and $0.9~\mathrm{K}$ would decrease the FRR from 92\% to 85.7\%, 74.6\%, and 63.2\%, respectively.

The median values of $T_{\mathrm{peak}}$ and angular areas are $~1.7~\mathrm{K}$ and $11.25~\mathrm{arcmin}^2$, respectively. 
There are 243 structures with $T_{\mathrm{peak}} \leq 3~\mathrm{K}$.
The remaining 57 structures with $T_{\mathrm{peak}}>3~\mathrm{K}$ hold more than $80\%$ of the total flux and create a tail between $\sim 3~\mathrm{K}$ and $\sim 15~\mathrm{K}$ in the $T_{\mathrm{peak}}$ distribution.
The bright subset with $T_{\mathrm{peak}}>3~\mathrm{K}$ has a flat angular size distribution, as shown in Fig.~\ref{fig:struct-prop-distrib}(b), while the distribution of all structures is skewed toward the method threshold ($N_{\mathrm{pix,min}}=16 \sim 4~\mathrm{arcmin}^2$).
As shown in Fig.~\ref{fig:struct-prop-distrib}(e), 18 structures have angular areas greater than $100~\mathrm{arcmin^{2}}$, covering $\sim 57\%$ of the total angular area of all structures. 

The velocity dispersion distribution of the bright subset is very similar to that of all structures, with median values of $1.14~\mathrm{km~s^{-1}}$ in the bright subset and $1.11~\mathrm{km~s^{-1}}$ among all structures.
Fig.~\ref{fig:struct-prop-distrib}(f) shows the velocity dispersion cumulative distributions of the bright and dimmer subsets, which seem identical. 
The p-value from the two-sample Kolmogorov-Smirnov test is 0.52, not rejecting the null hypothesis of identical distributions. 
In molecular clouds, the velocity dispersion is correlated to its physical scale, known as Larson's first relation \citep{1981MNRAS.194..809L}.
The detailed analysis requires accurate distance estimations, which are beyond the scope of this work.

\begin{figure*}[t!]
    \centering
    \includegraphics[width=\textwidth]{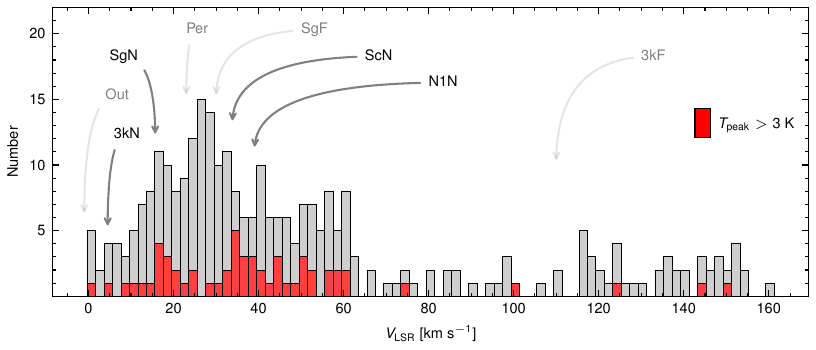}
    \caption{
    The centroid velocity ($\vlsr$) distribution of the three hundred structures. 
    The approximate radial velocities of spiral arm segments \citep{2019ApJ...885..131R} in this direction ($l=14^{\circ}$) are annotated. 
    Lighter gray arrows and labels denote the arm segments at the far side of the galaxy.
    Abbreviations of the arm segments include: SgN/F (Sagittarius arm Near/Far), ScN (Scutum arm Near), N1N (Norma arm 1st Quadrant Near), 3kN/F (3 kpc arm Near/Far), Per (Perseus arm), and Out (Outer arm).
    Structures with $T_{\mathrm{peak}} > 3~\mathrm{K}$ are shown in red. 
    }
    \label{fig:arms}
\end{figure*}

The data applied here is toward the inner galaxy at $l=14^{\circ}, b=0^{\circ}$, which means multiple spiral arms might be seen through this line of sight.
We plot the $V_{\mathrm{LSR}}$ distribution with the typical radial velocities of the spiral arms in Fig.~\ref{fig:arms}.
The approximate velocities of these arm segments are taken from \cite{2019ApJ...885..131R}.
We label the spiral arm segments at the near/far side with arrows in darker/lighter gray. 
Even though there seem to be some associations between the distribution peaks and the spiral arm velocities in the range of $[10, 50]~\mathrm{km~s^{-1}}$, the number of structures ($300$) is not large enough to have a solid determination of the velocity peaks. 
Therefore, a larger data set should be used for a more prominent correlation between the structure velocity and the spiral arm segments.

\section{Discussion}
\label{sect:discuss}
\subsection{Important Features of ISMGCC}
The proposed method has some features worth discussing.
Few assumptions are included in defining the molecular gas structures targeted by our method. 
In spatial dimensions, there is no prior assumption on the geometry of structures. 
The spatial boundary of each structure can have arbitrary geometry based on pixel connectivity.
Besides, we have no assumption on the number of Gaussian components on each pixel of a structure, which is different from the ACORNS \citep{2019MNRAS.485.2457H} algorithm where two components from the same location can never be linked to the same cluster.
We discard the single-component assumption because of the complicated line profile of \CO{13}{}{} emissions at high column density regions where the optical depth could be high and self-absorption might exist.
Fitting the complicated line profiles caused by these factors requires multiple Gaussian components.

There is always a compromise between isolating the blended structures and preserving more flux. 
Our method is designed to strike a balance between distinguishing blended structures and retaining as much flux as possible. 
The traditional single-linkage methods can distinguish multiple structures at the cost of increasing intensity cutoff values.
However, setting a high cutoff threshold would significantly reduce the FRR.
Many voxel-based methods use a signal-to-noise cutoff parameter to distinguish noise and signal.
But it is impossible to determine a clipping level at which most of the emission is kept and most of the noise is suppressed \citep{2011arXiv1101.1499D}.
The optimized parameter setting of our method given in Sect.~\ref{sect:metrics} has $\eta_0 = 0$, which means there is no global cutoff on the decomposed Gaussian components. 
This makes our method sensitive to the weak signals extracted by Gaussian decomposition.

\subsection{Connections with Previous Methods}
The ISMGCC method is not the first one that interprets PPV data with graphs.
An earlier graph-based method named SCIMES \citep[Spectral Clustering for Interstellar Molecular Emission Segmentation,][]{2015MNRAS.454.2067C} uses dendrograms as inputs and recasts them as graphs.
It regards the dendrogram leaves as the graph nodes (vertices) and the highest branches connecting node pairs as the edges, where the edge weights are defined through the similarity criteria based on the volume and/or luminosity of the dendrogram leaves. 
Then the graphs are optimally cut into segments through spectral clustering. 
Because the definitions of volume and luminosity consider each leaf's size (with distance) and velocity dispersion, SCIMES is more like a physical-oriented approach than a simple pixel/voxel-based segmentation method. 
The idea of using graphs to represent PPV data from SCIMES inspired the design of ISMGCC.
Even though both SCIMES and ISMGCC utilize weighted, undirected, and simple graphs, the graphs used in SCIMES are fully connected, i.e., there is always an edge between any pair of nodes in the graph, while those in ISMGCC are not.
The edges could only exist between Gaussian components with spatial separations smaller than the radii ($R_{\mathrm{ap}}$) in the aperture clustering procedure (Sect.~\ref{sect:aperture_clustering}) of ISMGCC. 
Therefore, instead of spectral clustering that works well on fully connected graphs, we have chosen a community-finding algorithm to cut the graphs.

Another important physical-oriented approach is the gravity-based G-virial method \citep{2015A&A...578A..97L}.
It generalized the virial parameter \citep{1992ApJ...395..140B} from the self-gravity of regions to the gravitational interactions between voxels in the PPV data cube. 
The G-virial value of a given voxel is the sum of gravitational boundedness from all the other voxels \citep[see Eqs. 5 and 6 in][]{2015A&A...578A..97L}. 
Even though it is still voxel-based, the definition of G-virial has considered the inequivalence between the spatial and velocity dimensions.
Because the gravitational boundedness ($I$) between two voxels is proportional to their spatial ($\delta_r$) and velocity  separations ($\delta_v$) with different indices, i.e., $I \propto \delta_r^{-1} \delta_v^{-2}$. 
This fact makes the irrelevant velocity components have minimal contributions to the G-virial of the current voxel.
Therefore, converting the PPV data cube into a G-virial cube and then finding the structures with DENDROGRAM or other methods could have a robust region definition. 
Besides, the recovered data cube for each structure identified by ISMGCC could also be the input of the G-virial method.
In this way, the gravitational coherent regions can be identified without the contamination from irrelevant velocity components.

\begin{figure*}
    \centering
    \includegraphics[width=1.0\linewidth]{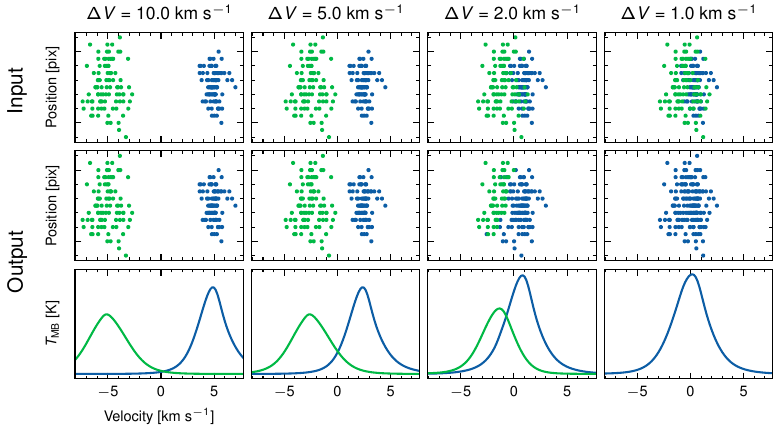}
    \caption{
    A simple simulation of ISMGCC results with various velocity crowdedness. 
    The top row contains the ``true'' structure finding results, where the blue and green points represent the Gaussian components in two structures. 
    These components are shifted along the velocity axis to various degrees and fed into ISMGCC again without their old affiliations. 
    The outputs in the middle row are the new structure finding results, while the bottom row displays the corresponding recovered average spectra. 
    The $\Delta V$ atop each column denotes the centroid velocity difference between the original structures.
    When $\Delta V$ is too small, e.g., the rightmost column, the two structures are not distinguishable and end as one structure. 
    }
    \label{fig:velocity-limit}
\end{figure*}

\subsection{Limitation of ISMGCC}
The design of ISMGCC absorbed some ideas from previous works. Meanwhile, its major difference from the voxel-based approaches is that the flux of each voxel can be assigned to multiple structures. 
This characteristic came from the Gaussian decomposition, through which ISMGCC could distinguish structures in crowded regions. 
However, it implicitly assumes that the spatially overlapped structures have distinguishable radial velocities, leaving an inevitable limitation on ISMGCC and any other methods working with PPV data. 
When the velocity difference between two overlapped structures is too small, they may be identified as one structure. 
This effect could be severe toward the regions with a shallower distance-velocity slope \citep[see fig. 7 in ][]{2022ApJ...925..201P}. 
We use two structures in our sample to demonstrate such a limitation.
To simulate various velocity crowdedness, we shift their Gaussian components with four centroid velocity differences ($\Delta V$) between them. 
Their Gaussian components are also relocated until the two structures are spatially overlapped.
Figure~\ref{fig:velocity-limit} demonstrates the ISMGCC results with the four $\Delta V$ values $[10, 5, 2, 1]~\mathrm{km~s^{-1}}$, from left to right, respectively. 
As $\Delta V$ decreases, the two structures gradually become indistinguishable.  
This simple simulation indicates that the ISMGCC method could fail to distinguish the structures blended on both the spatial and velocity axes.

\section{Summary}
\label{sect:summary}

In this work, we propose a new method named ISMGCC to find molecular gas structures in emission line data cubes. 
Based on Gaussian decomposition and graph theory, our method could disentangle the complicated molecular gas distributions in crowded regions with minimal flux loss and no prior assumptions of the structure geometry.
It also does not limit the number of Gaussian components on each pixel of a structure, enabling it to tolerate distorted line profiles.
All parameters are either dimensionless or in units of pixels, reducing the difficulty for users in adjusting the parameters on different datasets. 

We tested the ISMGCC method on the MWISP \CO{13}{}{} data in the range of $13.5^{\circ} \leq l\leq 14.5^{\circ}, |b|\leq 0.5^{\circ}$, and $-100\leq \vlsr \leq 200~\mathrm{km~s^{-1}}$. The main results  are as follows:
\begin{enumerate}
    \item Three hundred structures with at least 16 pixels are identified;
    \item These structures retain $92\%$ of the flux in the raw data cube and have single-peaked line profiles on $93\%$ of their pixels;
    \item The less numerous large-scale structures dominate the total flux and angular area;
    \item Unlike the flux and angular area, the velocity dispersion seems less affected by $T_{\mathrm{peak}}$. Structures above and below $T_{\mathrm{peak}} = 3~\mathrm{K}$ have identical distributions on their velocity dispersion;
    \item The centroid velocity distribution is not significantly correlated with the expected spiral arm segments because of the small structure number. Data with wider spatial coverage might bring more insights on this.
\end{enumerate}

Some possible improvements to ISMGCC could be made in the future.
For example, using multiple emission lines with various optical depths, chemical abundance, and critical density simultaneously might help find more complete structures across large physical condition ranges.
As for the MWISP data, we could use \CO{}{18}{} data to trace the centroid velocity at the high-column-density regions and apply \CO{12}{}{} to determine the outskirts of the gas structure.
Furthermore, more multi-line surveys have been done in recent years, e.g., CHaMP \citep{2011ApJS..196...12B}, FUGIN \citep{2017PASJ...69...78U}, Mopra-CO \citep{2013PASA...30...44B}, ThrUMMS \citep{2015ApJ...812....6B}, FQS \citep{2020A&A...633A.147B}, SEDIGISM \citep{2017A&A...601A.124S}, and OGHReS \citep{2024MNRAS.528.4746U}.
These surveys have acquired data with various molecules and transitions, providing great application scenarios for our method. 
Using ISMGCC with the future multi-line improvement might bring us a more comprehensive picture of the gas structure in molecular ISM.

\normalem
\begin{acknowledgements}
This work makes use of the data from the Milky Way Imaging Scroll Painting (MWISP) project, which is a multi-line survey in \CO{12}{}{}/\CO{13}{}{}/\CO{}{18}{} along the northern galactic plane with PMO-13.7m telescope. 
We are grateful to all the members of the MWISP working group, particularly the staff members at the PMO-13.7m telescope, for their long-term support. 
MWISP is sponsored by the National Key R\&D Program of China with grants 2023YFA1608000, 2017YFA0402701, and the CAS Key Research Program of Frontier Sciences with grant QYZDJ-SSW-SLH047.
This work was supported by the National Natural Science Foundation of China (NSFC, Grant Nos. U2031202, 12373030, and 11873093).
Z.C. acknowledges the Natural Science Foundation of Jiangsu Province (grants no. BK20231509).
We thank Yu Jiang, Yuehui Ma, and Xiangyu Ou for the discussion and trial use of the code that help improve the paper. 
\end{acknowledgements}
  
\bibliographystyle{raa}
\bibliography{bibtex}

\end{document}